# Wavelength-dependent photothermal imaging probes nanoscale temperature differences among sub-diffraction coupled plasmonic nanorods

*Seyyed Ali Hosseini Jebeli[⊥|], Claire A. West[‡|], Stephen A. Lee[†], Harrison J. Goldwyn[‡], Connor R. Bilchak[&§], Zahra Fakhraai[&], Katherine A. Willets[¶\*], Stephan Link[†⊥\*], David J. Masiello[‡\*]*

[⊥]Department of Electrical and Computer Engineering, Rice University, Houston, TX 77005, USA

[‡]Department of Chemistry, University of Washington, Seattle, WA 98195, USA

[†]Department of Chemistry, Rice University, Houston, TX 77005, USA

[§] Department of Materials Science and Engineering, University of Pennsylvania, Philadelphia, Pennsylvania 19104, USA

[&]Department of Chemistry, University of Pennsylvania, Philadelphia, PA 19104, USA

[¶]Department of Chemistry, Temple University, Philadelphia, PA 19122, USA

**Abstract**

While the thermal and electromagnetic properties of plasmonic nanostructures are well understood, nanoscale thermometry still presents an experimental and theoretical challenge. Plasmonic structures can confine electromagnetic energy at the nanoscale, resulting in local, inhomogeneous, controllable heating. But reading out the temperature with nanoscale precision using optical techniques poses a difficult challenge. Here we report on the optical thermometry of individual gold nanorod trimers that exhibit multiple wavelength-dependent plasmon modes resulting in measurably different local temperature distributions. Specifically, we demonstrate how photothermal microscopy encodes different wavelength-dependent temperature profiles in the asymmetry of the photothermal image point spread function. These point spread function asymmetries are interpreted through companion numerical simulations of the photothermal images to reveal how differing thermal gradients within the nanorod trimer can be controlled by exciting its hybridized plasmonic modes. We also find that hybrid plasmon modes that are optically dark can be excited by our focused laser beam illumination geometry at certain beam positions, thereby providing an additional route to modify thermal profiles at the nanoscale beyond wide-field illumination. Taken together these findings demonstrate an all-optical thermometry technique to actively create and measure thermal gradients at the nanoscale below the diffraction limit.

Harnessing the thermal response of optically excited noble metal nanoparticles has been used for a variety of different applications including drug delivery,[1-3] photothermal therapies,[4-6] photocatalysis,[7-8] heat generation from solar energy,[9-10] heat assisted magnetic recording,[11-13] and thermal manipulation of materials at the nanoscale.[14] In all of these applications, light interacts strongly with the nanoparticles and drives coherent charge oscillations known as localized surface plasmon (LSP) resonances. It is through the nonradiative decay of the LSP that the nanoparticle is heated, which in turn leads to temperature increases in the environment.[15-17] Measuring and understanding the induced nanoscale temperature gradients is critical for the optimization of many photothermal applications.[18-21]

Current research in nanothermometry has focused on different methods to measure temperature at the nanoscale. For example, the ratio of Stokes to anti-Stokes photoluminescence emission spectra has been used to calculate nanoparticle temperatures; however, it requires the lattice and electron temperatures to be approximately the same to correctly interpret the results.[22-24] Electron energy-loss spectroscopy (EELS) uses phonon energy shifts combined with ratioing loss-to-gain signals to infer the lattice temperature of the material.[25-26] However, while EELS can map temperature changes at truly nanoscale dimension, it requires a high-fidelity electron microscope and ultrahigh vacuum conditions which adds to the complexity of experiments. Alternatively, photothermal microscopy is an optical, *in situ* method that relies on nanoparticle and local environment heating to collect signal.[27-29] This method uses two lasers: a heating laser tuned to excite the nanoparticle and induce nanoscale thermal gradients and a second laser to probe refractive-index changes in the system induced by these thermal modifications.[30] The probe wavelength can be chosen off-resonance to avoid melting the particles while having high power to increase the signal to noise ratio (SNR).[30-32] A previous demonstration of this method in individual plasmonic nanorod dimers showed that the centroid of the photothermal signal is biased towards the hotter nanorod within the hybridized dimer, indicating the sensitivity of the technique to spatially non-uniform thermal profiles.[33]

In this work, photothermal microscopy is performed on individual, asymmetric gold nanorod trimers and the resulting images are correlated with their wavelength-dependent nanoscale thermal profiles. The nanorod trimer structure is designed to host three unique nanoscale temperature distributions associated with its three hybrid LSP modes. In imaging these modes, we find that the full width at half maximum (FWHM) of the photothermal image in both *x*- and *y*-directions of the image plane (obtained by fitting a 2-dimensional Gaussian) varies with pump (heating beam) wavelength. We explain these trends through LSP mode analysis, coupled optical and heat diffusion simulation, and photothermal image modeling using a focused illumination source,[33-34] revealing their origin in the asymmetric excitation of the nanorod trimer's hybrid LSP modes and their associated thermal distributions. It is through this analysis of the ratio of *x*- and *y*-direction FWHM of the photothermal image that we uncover and herein report upon a new approach for all-optical thermometry at the nanoscale.

**Results and discussion**

Figure 1a displays the gold nanorod trimer designed to produce wavelength-dependent, nanolocalized temperature distributions under steady-state optical excitation (see complete simulation details in Supporting Information section S1). The nanorod trimer's three hybrid LSP modes, labeled $\lambda_1$, $\lambda_2$, and $\lambda_3$. The nanorods are each 80 nm long and 40 nm wide with a 20 nm separation, making the entire system (180 nm × 100 nm), well below the diffraction limit of the focused probe laser (~360 nm). Each nanorod trimer is nanofabricated on a glass substrate and immersed in glycerol to provide a low thermally conductive environment for heat to diffuse. Nineteen such nanorod trimers were nanofabricated using electron beam lithography (see complete experimental details in Supporting Information section S2). Absorption and dark-field scattering spectra of all studied individual nanorod trimers were measured, each having the same general features as the blue traces shown in Figure 1b. The absorption spectra are obtained using a photothermal microscope and the scattering spectra are measured using a hyperspectral microscope.[35-36] In both setups, the nanorod trimers are excited using a numerical aperture (NA) of 1.4 and the light is collected using a 0.7 NA objective after interacting with the structures. Simulated absorption and scattering cross-sections resulting from a focused Gaussian beam excitation source, to mimic the photothermal imaging experiments, using full-wave electromagnetic simulations in a modified discrete dipole approximation[37-38] are shown in the orange traces in Figure 1b. The beam centroid lies at the center of the nanorod trimer and well approximates the experimental spectrum collected at the same point. The two resonances in the simulated absorption spectrum correspond to the hybrid LSP modes $\lambda_1$ and $\lambda_2$. These bright modes are indistinguishable in our measurements due to the broad linewidth of the laser (10 nm determined by the acousto-optic filter that selects the pump wavelength from a white-light laser), while the third hybrid mode, $\lambda_3$, is inaccessible when the pump beam is focused at the nanorod trimer center. Due to small imperfections in the otherwise equivalent nanorod trimers, each of the 19 structures has a slightly different absorption spectrum. Figure 1c shows a histogram of the wavelength corresponding to the peak in their individual absorption spectra.

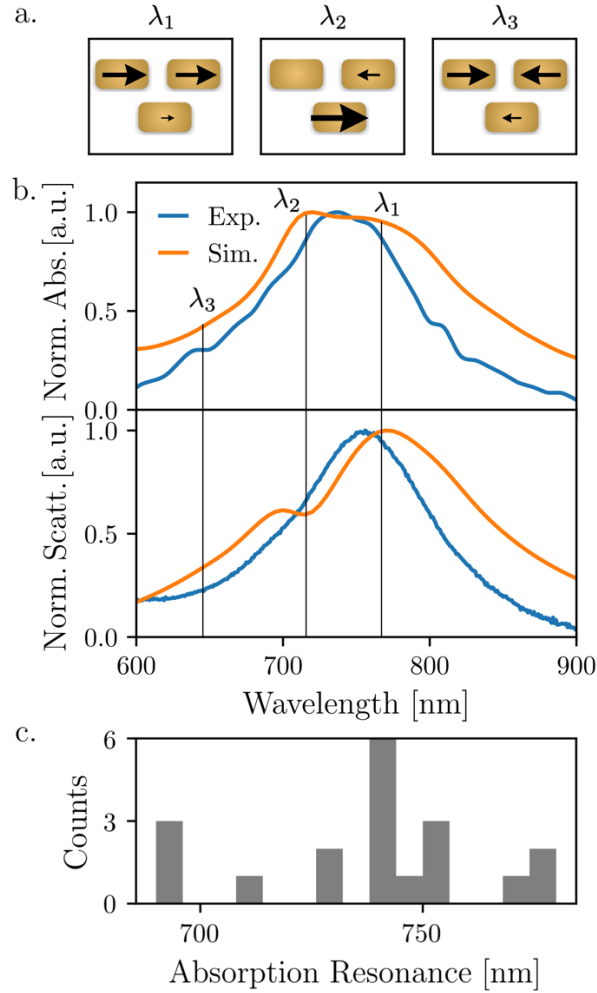

Figure 1. Absorption and scattering spectra of the individual, nanofabricated and simulated nanorod trimer structures. a) Hybrid dipolar modes of the nanorod trimer are calculated using a coupled oscillator model. b) Absorption spectrum (top) of one of the nanorod trimers measured with photothermal spectroscopy using a 532 nm probe laser centered on the nanorod trimer (blue line). The simulated absorption spectrum (orange line) resolves two modes, labeled $\lambda_1$ and $\lambda_2$. The highest energy mode, $\lambda_3$, is inaccessible at this beam position. Scattering spectra (bottom) of the same nanorod trimer measured (blue line) and computed (orange line) are shown. The resonance at 700 nm is not resolved in experiment likely due to nanofabrication

differences between simulation and experiment. c) A histogram depicting the variability in experimental absorption resonance position of 19 individual nanorod trimers studied.

Our photothermal microscopy approach relies on raster scanning the beam within a 1 μm x 1 μm region of interest surrounding an individual structure. Even though the nanorod trimer's dimensions are well below the diffraction limit, different centroid positions of the focused pump beam drive different weightings of the three hybrid LSP modes. Figure 2 depicts the simulated dependence of the absorption spectrum on the position of the pump beam centroid. The schematic in Figure 2a labels the location of five different focused beam positions where absorption spectra are calculated. The schematic (top left) depicts the pump beam waist when centered at the origin for wavelengths ranging from 600 nm (dark gray) to 900 nm (lighter gray).

The resulting absorption spectra as a function of beam position are shown in Figure 2b where the colors and line styles of the spectra correspond to the illustration in Figure 2a. When the beam is positioned to the left of the nanorod trimer (solid red ⊕), above the nanorod trimer (dashed blue ⊕), and in the center of the nanorod trimer (solid black ⊕), the two lowest energy modes at $\lambda_1$ and $\lambda_2$ are approximately driven the same. The third mode is nearly completely undriven. However, when the beam is positioned beneath the nanorod trimer (dashed red ⊕), the $\lambda_2$ mode is driven most strongly. Lastly, when the beam is positioned to the right of the nanorod trimer (solid blue ⊕), the previously plane-wave dark $\lambda_3$ mode can be driven nearly as strongly as the other two modes. The pump wavelength specific absorption cross-section maps in Figure 2c depict this absorption asymmetry at the three LSP mode wavelengths as a function of beam centroid position. The black contour lines represent equipotentials of constant absorption.

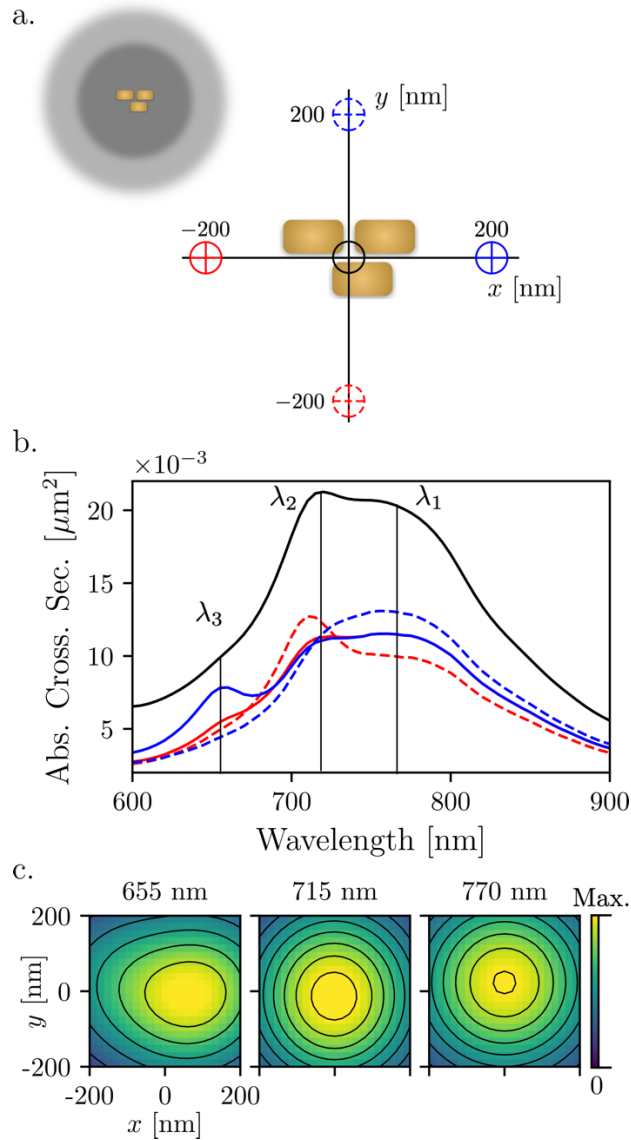

Figure 2. The focused pump beam drives different weightings of the three hybrid LSP modes depending on its centroid position relative to the nanorod trimer structure. a) An illustration of the five beam positions with respect to the position of the nanorod trimer selected for simulations. The colors and line styles of the cross hairs correspond to the absorption spectra in the following panel. The schematic in the top left corner indicates the beam waist size when at the origin for a 600 nm wavelength pump beam (dark gray) and 900 nm wavelength pump beam (light gray). b) Different absorption spectra resulting from the focused beam driving the nanorod trimer at the five labeled beam positions. When the beam is positioned at the solid red ⊕, the dashed blue ⊕, and the solid black ⊕, the two lowest energy modes are driven approximately equally. However, when the beam is located at the dashed red ⊕ or the solid blue ⊕, the $\lambda_2$ or $\lambda_3$ modes are strongly driven, respectively. c) Absorption cross-section maps as a function of beam centroid at the three resonances. The overlaid black lines show equal contours of the absorption map. The color scale of the 655 nm image has been scaled up by a factor of two for ease in comparison with the more intense absorption maps at 715 nm and 770 nm.

Each of these absorption profiles underlies different temperature distributions that evolve as a function of pump beam position and wavelength. Figure 3 depicts the computed thermal profiles associated with the five beam centroid positions displayed in Figure 2. Here, the steady-state temperature profiles are calculated using the thermal discrete dipole approximation,[39] and reflect the temperature of the particles above ambient room temperature. When the structure is probed at the lowest energy LSP mode, $\lambda_1 = 770$ nm, the top two particles heat up more than the bottom particle at every beam position in the 400 nm wide window shown. At the second hybrid mode, $\lambda_2 = 715$ nm, the temperature pattern switches with the bottom particle reaching higher temperatures than the top two for most (although not all) beam positions. Finally, the last hybrid mode, $\lambda_3 = 655$ nm, produces a more complex thermal profile that depends on the pump beam position with absolute and relative nanorod temperatures smaller than those found at $\lambda_1$ and $\lambda_2$. See Figure S1 for temperature difference maps at each beam position.

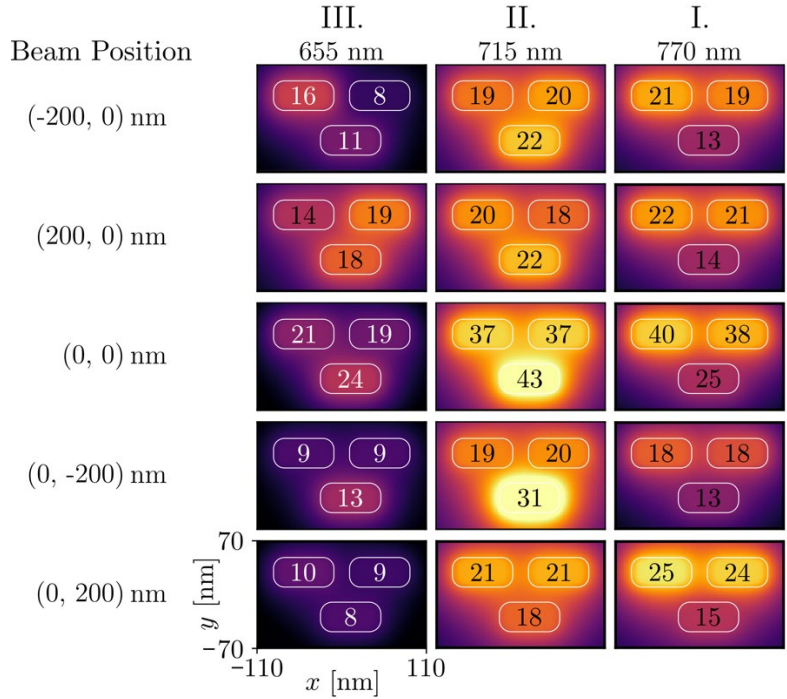

Figure 3. Wavelength and beam centroid dependent computed temperature maps of the nanorod trimer. Steady-state temperature maps of the nanorod trimer at the three hybrid modes and five unique beam positions labeled in Figure 2a. The temperatures listed are degrees Celsius above room temperature.

The prediction that the induced temperature shifts between the top region being hotter (when excited around 770 nm) and the bottom region hotter (around 715 nm) is probed with photothermal microscopy. The images of 19 nanorod trimers were acquired by raster scanning the sample across the pump and probe lasers in a collinear focused beam geometry. 180 nm diameter gold nanodisks were fabricated alongside each nanorod trimer and are expected to generate uniform heating, thereby acting as a reference. Figure 4 shows the resulting photothermal images at two pump wavelengths, 700 nm and 800 nm, near the two lowest energy LSP modes of a representative nanorod trimer and nanodisk pair. The differences among the images are quantified by fitting the data to a point spread function (PSF) described by a 2-dimensional Gaussian, $I(x,y) = A \cdot \exp\left(-\frac{(x-x_0)^2}{2\sigma_x^2} - \frac{(y-y_0)^2}{2\sigma_y^2}\right)$ where $(x_0, y_0)$ is the beam centroid, $A$ is the amplitude, and $\sigma_x, \sigma_y$ are the Gaussian widths in $x$ and $y$. The Gaussian widths were converted into FWHM by the relationship $\text{FWHM}_{x,y} = 2\sqrt{2 \ln 2} \cdot \sigma_{x,y}$. The fits are shown in the second row of Figure 4. As expected, $\text{FWHM}_x = FWHM_y$ for the PSFs of the nanodisks under both 700 nm and 800 nm pump wavelengths. This result serves as a reference allowing us to correct for any wavelength-dependent asymmetry due to variation within our imaging setup. The nanorod trimer on the other hand exhibits a different PSF than the nanodisk. The nanorod trimer image at 700 nm is elongated along the $x$-direction resulting in an asymmetry factor ($\text{FWHM}_x/FWHM_y$) larger than one. At 800 nm, the nanorod trimer image appears symmetric.

Simulated photothermal images are used to interpret the observed asymmetries in the experimental nanorod trimer PSFs. The simulated images are obtained by raster scanning a focused Gaussian beam across a nanorod trimer or nanodisk on glass and integrating the scattered field over the solid angle $(\theta_d, \phi_d)$ spanning the detector in the forward direction according to[40-41]

$$I^{\text{PT}} = \frac{cn}{8\pi} \iint\limits_{(\theta_d, \phi_d)} \left(|\mathbf{E}_H(\theta, \phi)|^2 - |\mathbf{E}_R(\theta, \phi)|^2\right) d\Omega$$

where $c$ is the speed of light, $n$ is the refractive index at the detector, $\mathbf{E}_{H,R}(\theta, \phi)$ is the probe electric field scattered through the heated ($H$) or room ($R$) temperature system, and $\theta, \phi$ are evaluated at the detector located 1 cm away. The collection angle of the photothermal experiment defines the bounds of integration for the simulation, which are set to be $\theta_d = 35°$ and $0° \leq \phi_d \leq 360°$. The scattered electric field through the heated system, $\mathbf{E}_H(\theta, \phi)$, is calculated by assigning the gold refractive index values according to the temperatures calculated at each wavelength and beam position. See Supporting Information section S4 for a description of the experimental temperature-dependent refractive indices of gold and section S3 for comparison to an alternative image function derived in Ref. 42. The infinite glycerol background is assigned an increased refractive index corresponding to the average temperature of the gold particle(s). The resulting simulated photothermal images of the nanorod trimer and nanodisk are shown in the last row of Figure 4. The simulated PSFs for the nanodisk are symmetric for both the 700 nm and 800 nm pump, as was the case for the experimental images. The simulated nanorod trimer PSFs exhibit nearly symmetric PSFs at both 700 nm (FWHM$_x$/FWHM$_y$ = 0.99) and 800 nm pump (FWHM$_x$/FWHM$_y$ = 1.02), differing from the experimental results. This apparent contradiction between theory and experiment is further explored by comparing the spectrally resolved simulations and experimental distributions.

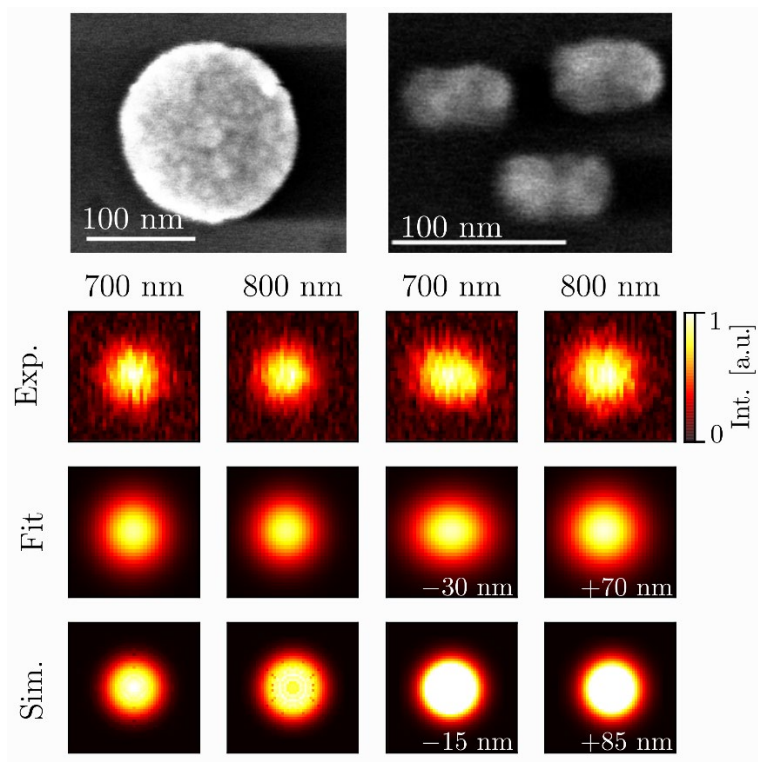

Figure 4. Photothermal images of a nanodisk (left two columns) and a nanorod trimer (right two columns) with representative SEM images shown on top. The image sizes of the experimental, fit, and simulated images are 1 μm × 1 μm. The first row depicts experimentally obtained photothermal images of the structures pumped at 700 nm and 800 nm with 190 μW and 170 μW powers. The second row shows the fit of the raw data to a 2-dimensional Gaussian function. The asymmetry in the PSF of the nanorod trimer at 700 nm is apparent and juxtaposed against the symmetric nanodisk images. The last row depicts the simulated photothermal images of the nanodisk and nanorod trimer. The printed values on the fit and simulated nanorod trimer images indicate the difference in wavelength between the pump wavelength and the nanorod trimer's absorption maximum. Each image has been individually normalized.

Statistics of the PSF fit parameters extracted from the photothermal images are shown in Figure 5 for the 19 nanorod trimers investigated. The values have been corrected to account for effects of alignment drift and aberrations. The correction involves two steps: first, the PSF FWHM in each direction is normalized by the value of the nanodisk in the same direction (e.g., $\text{FWHM}_{x_{\text{trimer}}}/\text{FWHM}_{x_{\text{disk}}}$). The nanodisks and nanorod trimers are fabricated next to each other on the sample and a nanodisk-nanorod trimer pair is always imaged together, thus drift and aberrations in the system are captured by the nanodisk PSF and corrected for in the nanorod trimer PSF. In the second step the normalized nanorod trimer FWHM values are multiplied by the average nanodisk FWHM as a universal scaling factor. Note that even without normalization and correction we still observe the same signature in the nanorod trimer images (Figure S2).

Figure 5a presents a histogram of the FWHM in the *x*- and *y*-directions from the fits of the nanodisk and nanorod trimer images at the two pump wavelengths. Just as qualitatively observed in Figure 4, under 700 nm pump, the FWHM in the *x*-direction (blue) is larger than the FWHM in the *y*-direction (orange) for nearly all nanorod trimers. Similarly, under 800 nm pump the FWHM in the *x*-direction is larger than the FWHM in the *y*-direction, but to a smaller extent. The statistical test results summarized in Table S1 confirm a statistically significant difference between FWHM$_x$ and FWHM$_y$ at 700 nm. Similarly, a smaller yet statistically significant difference at 800 nm is also observed.

To rule out any effect of scan direction or pump beam polarization, the sample was imaged in two different scan directions and two sample directions (Figure S3). The results in Figure S4 show that scan direction has no effect on the values of the nanodisk-corrected asymmetry factor at 700 nm pump. Also, any effect possibly resulting from sample orientation at 700 nm pump is negligible as shown in Figure S5. The correlation plot in Figure S6 illustrates that there is no correlation in scan direction at 800 nm excitation as seen from the data clustering around (1, 1). Furthermore, the asymmetry factors at the two different pump wavelengths are negatively correlated (Figure S7). These control experiments verify that the observed PSF asymmetries are generated by changing the wavelength of the pump beam and not by sample orientation or scan direction.

To understand the full wavelength-dependence of the photothermal images of the nanorod trimer, the data from Figure 5a is replotted to show the FWHM ratio (FWHM$_x$/FWHM$_y$) at the 700 nm pump wavelength (green dots) and at the 800 nm pump wavelength (purple dots) (Figure 5b). The experimental and simulated data are plotted on a common *x* axis by subtracting the pump wavelength from the nanorod trimer absorption maximum to find the distance of the resonance maximum from the pump wavelength ($\Delta\lambda$). Specifically, each nanorod trimer has a different absorption maximum with a range spanning from 690 nm to 780 nm (Figure 1c). By calculating the differences between the pump wavelengths of 700 nm and 800 nm and the resonance maxima of the nanorod trimer, a range of $\Delta\lambda$ values between -80 nm and +110 nm is obtained. Simulated FWHM ratios at different pump wavelengths for a fixed calculated absorption spectrum (Figure 1b) are overlaid and given by the black line. These simulated FWHM ratios are obtained by fitting the photothermal images of the nanorod trimer pumped with a wavelength range of 610 nm to 790 nm. The maximum in the simulated absorption spectrum of the nanorod trimer (715 nm) is then subtracted from the pump wavelength range.

When the pump beam excites the nanorod trimer at wavelengths greater than or equal to the $\lambda_1$ ($\Delta\lambda \geq 55$ nm) absorption resonance, the FWHM ratio is greater than one (Figure 5b). This excitation corresponds to the lowest energy hybrid mode, where the upper two nanorods heat up more than the lower nanorod (Figure 3 Column I). The thermal lens created by the hotter upper nanorod pair is broader in the *x* direction than the *y* direction leading to an elongation in the photothermal image in the *x*-direction and an asymmetric PSF. As the excitation wavelength approaches the $\lambda_2$ absorption resonance and goes slightly to shorter wavelengths ($\Delta\lambda \approx 0$ nm), the FWHM ratio then dips below one (Figure 5b). In this case, the lower nanorod for most beam positions is significantly hotter than the upper pair (Figure 3 Column II) and the associated image distorts to favor the *y*-direction. Finally, as the pump beam excites the nanorod trimer at shorter wavelengths near the $\lambda_3$ absorption resonance ($\Delta\lambda = -60$ nm), the plane-wave dark mode becomes optically accessible to the focused beam excitation with a highly spatially-dependent thermal profile. The excitation of the dark mode results in the increased asymmetry observed at $\Delta\lambda = -60$ nm, which occurs when the pump wavelength is blue-shifted 60 nm from the absorption peak of the nanorod trimer.

The PSF asymmetries induced by beam position and pump wavelength are made more apparent when comparing the difference in absorption cross-sections at the spatial positions marked in Figure 2a. Figure 5c demonstrates that the pump wavelength-dependent asymmetry is qualitatively reproduced by the differences in the spatially-dependent absorption cross-sections shown in Figure 2b calculated at different focused illumination beam positions. Those absorption cross-sections have been reproduced in the insets of Figure 5c. Specifically, we find that by taking the difference between the absorption cross-sections at (-200 nm, 0 nm; solid red) and (0 nm, -200 nm; dashed red), the resulting trace (dot-dashed red) changes sign where the FWHM ratio (black) dips above and below one at nearly the same wavelengths. Further insight is gained when similarly taking the difference between the absorption cross-sections at (200 nm, 0 nm; solid blue) and (0 nm, 200 nm; dashed blue). In this case, there is a maximum at the same position as the dominant peak in the ratio plot. This maximum is due to the asymmetric driving capability of the focused beam in exciting the plane-wave dark mode $\lambda_3$

Certain beam positions preferentially excite one hybrid mode more strongly than the others, contributing to the distortion of the photothermal image. This result makes clear how the peak in the FWHM ratio near 650 nm ($\Delta\lambda = -60$ nm) is due to the pump beam asymmetrically exciting the third hybrid mode.

Except for the $\lambda_3$ mode in the lower panel of Figure 5c, we observe that these cross-section differences change sign (indicated by the shaded spectral regions) at approximately the same wavelengths where asymmetries in the PSFs occur, as reflected in the plotted FWHM ratio. Thus, these trends reveal how the measured asymmetry in the FWHM originates from spatial differences in absorption that depend upon the location of the pump beam. The observed PSF asymmetries are therefore correlated with asymmetric local heating of the nanorod trimer and the resulting inhomogeneous temperature changes that ensue. Taken together, these trends in the FWHM ratio reflect the spatial and wavelength-dependence of the absorbing hybrid LSP modes as well as their associated spatially- and wavelength-dependent thermal responses. Therefore, by combining our photothermal imaging measurements with corresponding numerical simulations, we have discovered an indirect route to retrieve nanoscale thermal information using purely optical techniques.

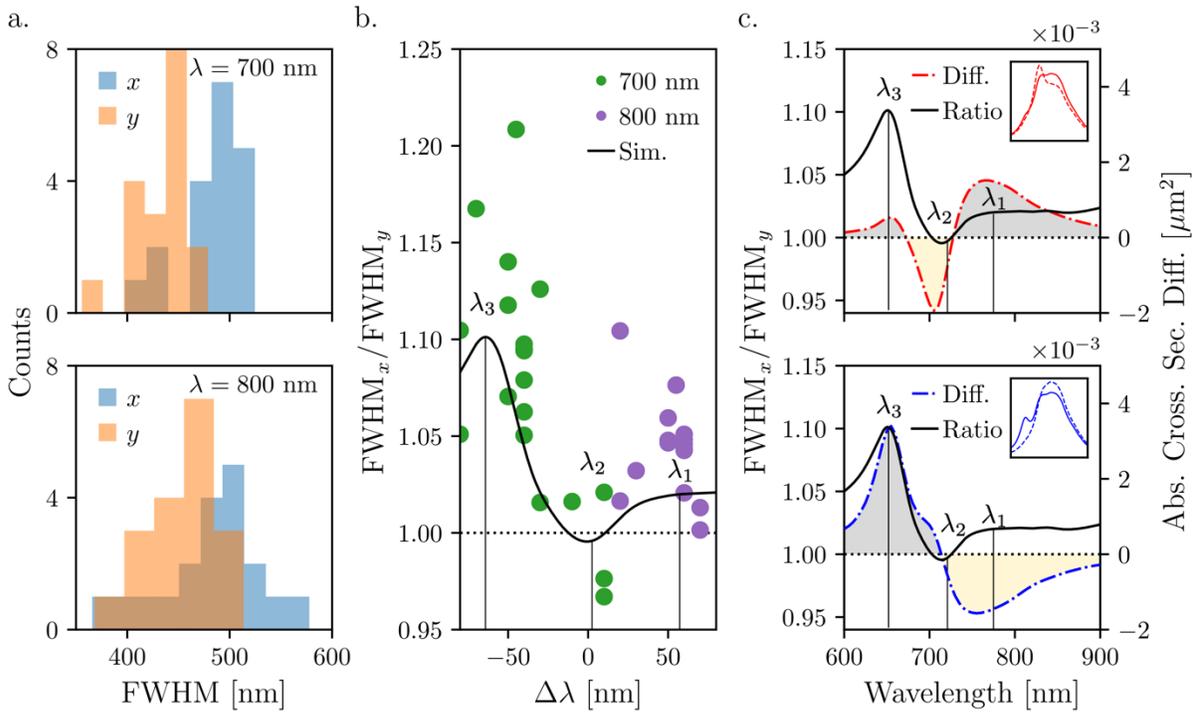

Figure 5. Spectra and photothermal image fitting results of different nanorod trimer structures. a) Values of the FWHM in the *x*- and *y*-directions corrected by the nanodisk at 700 nm and 800 nm pump wavelengths are shown. b) Experimental *x* and *y* FWHM ratios at 700 nm (green dots) and 800 nm (purple dots). The simulated FWHM ratios are shown by the black line. To align the experimental and simulated data on the same *x*-axis, the difference between the peak

absorption and the pump wavelength is used. The distribution in the asymmetry values comes from the different absorption maxima of the 19 nanofabricated nanorod trimers (Figure 1c). c) Differences between the absorption cross-sections (from Figure 2b and reproduced in the insets) as a function of beam position. Absorption cross-section differences (dot-dashed curves) between beam positions (-200 nm, 0 nm) and (0 nm, -200 nm) indicated in red (upper panel) and beam positions (200 nm, 0 nm) and (0 nm, 200 nm) indicated in blue (lower panel).

**Conclusion**

In conclusion, we have demonstrated how asymmetries in the photothermal images of individual gold nanorod trimers excited at their hybridized LSP resonances are correlated with the spatially inhomogeneous thermal profiles associated with each mode. In particular, we show how sub-diffraction-limited thermal gradients within the nanorod trimer can be directed using focused laser excitation at different wavelengths and how these gradients are encoded as asymmetries in the FWHM of the 2-dimensional PSFs obtained from the photothermal images. Theoretical modeling of the nanorod trimer's optical and thermal responses together with the imaging optics allows us to explicitly connect the experimental photothermal images with precise nanoscale temperature values. In this manner, we have demonstrated a new procedure—combining experimental imaging and photothermal modeling—to perform all-optical thermometry measurements on individual nanoscale objects that are smaller than the diffraction limit of light.

ASSOCIATED CONTENT

**Supporting Information**.

Simulated spectra and imaging details, experimental spectra and imaging details, temperature differences at different beam positions; temperature dependent ellipsometry of gold; tabulated temperature dependent Drude-Lorentz parameters of gold; PSF FWHM distributions; correlations plots of PSF width under different imaging conditions (PDF).

AUTHOR INFORMATION

Corresponding Authors

Katherine A. Willets, kwillets@temple.edu; Stephan Link, slink@rice.edu; David J. Masiello, masiello@uw.edu

Author Contributions

The manuscript was written through contributions of all authors. All authors have given approval to the final version of the manuscript. ‖These authors contributed equally.

ACKNOWLEDGMENTS

This work was supported by the U.S. National Science Foundation under grant nos. NSF CHE-1727092 (D.J.M), CHE-1727122 (S.L.), and CHE-1728340 (K.A.W.). S.L. also acknowledges support from the Robert A. Welch Foundation (grant no. C-1664). The experimental work was conducted in part using resources of the Shared Equipment Authority at Rice University. The theoretical work was facilitated through the use of advanced computational, storage, and networking infrastructure provided by the Hyak supercomputer system at the University of Washington.

ABBREVIATIONS

PSF, point spread function; SNR, signal to noise ratio; FWHM, full width at half maximum; NA, numerical aperture.

For Table of Contents Only

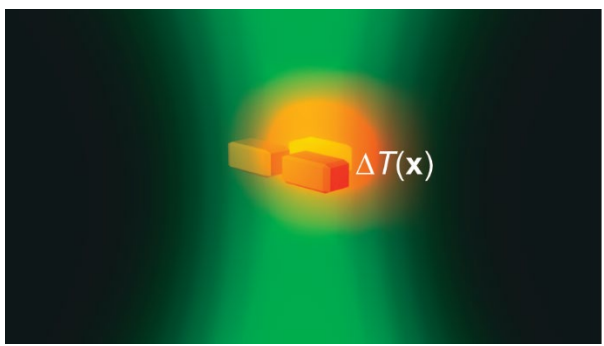

# Supporting Information: Wavelength-dependent photothermal imaging probes nanoscale temperature differences among sub-diffraction coupled plasmonic nanorods

*Seyyed Ali Hosseini Jebeli[⊥|], Claire A. West[‡|], Stephen A. Lee[†], Harrison J. Goldwyn[‡], Connor R. Bilchak[&§], Zahra Fakhraai[&], Katherine A. Willets[¶\*], Stephan Link[†⊥\*], David J. Masiello[‡\*]*

[⊥]Department of Electrical and Computer Engineering, Rice University, Houston, TX 77005, USA

[‡]Department of Chemistry, University of Washington, Seattle, WA 98195, USA

[†]Department of Chemistry, Rice University, Houston, TX 77005, USA

[&]Department of Chemistry, University of Pennsylvania, Philadelphia, PA 19104, USA

[§] Department of Materials Science and Engineering, University of Pennsylvania, Philadelphia, Pennsylvania 19104, USA

[¶]Department of Chemistry, Temple University, Philadelphia, PA 19122, USA

[|]These authors contributed equally.

**Table of Contents, Figures, and Tables**                                                                                      Page

**Section S1: Simulation Methods**

Coupled optical, heat diffusion, and image modeling

The computational approach for obtaining the photothermal images used DDSCAT 7.3[1] modified to account for a Gaussian beam excitation source and the thermal discrete dipole approximation (T-DDA).[2] The scattering calculations were performed on gold nanorod trimers using an experimentally obtained temperature-dependent dielectric data set (See S1) with 2 nm dipole spacing in a uniform infinite glycerol background with refractive index $n = 1.473$. The resulting electric field and polarization information within the nanorods were used as inputs for the steady-state temperature calculation performed in T-DDA, which includes a semi-infinite silica ($\kappa = 1.38$ W/m K) substrate. The temperatures of each nanorod were then used to select the appropriate temperature-dependent dielectric values for a second scattering calculation. This second scattering calculation was performed at an off- resonant probe wavelength of 600 nm with the nanoparticles embedded in a uniform background with a temperature-dependent refractive index value evaluated at the average nanoparticle temperature of the trimer. The resulting scattered electric fields were evaluated on a partial hemisphere 1 cm away from the particles and were used to calculate the total intensity as shown in Eq. 1. To obtain the entire two dimensional photothermal image, this full procedure was carried out for a range of co-scanned excitation and probe beam positions that spanned a region of interest surrounding the nanorod trimer system.

**Section S2: Experimental Details**

Sample preparation:

The samples were prepared using electron beam lithography on glass slides. The slides were cleaned by sonicating for 15 min in a base piranha solution ($NH_4OH:H_2O_2:H_2O$ 1:4:20) followed by rinsing and sonication in deionized H₂O. The slides were finally treated with oxygen plasma for 1 min. PMMA950 (MicroChem A2 solution) was then spin coated onto a cleaned slide at 3000 rpm for 1 min and a conductive polymer layer (E spacer, Showa Denko) was added by spin coating at 3000 rpm for 1 min to dissipate charge while writing the pattern. The pattern writing was performed on a JEOL electron microscope at 30 kV with 40 pA current and an area dose of 400 μC/cm$^2$ at 4 nm line spacing. After writing the pattern, the samples were rinsed with water for 20 seconds to remove the E-spacer. The pattern was developed in an isopropyl alcohol: water 7:3 solution for 65 seconds followed by water for 35 seconds. This method ensures accurate development with more control over the gap size, which is an important factor for the trimer structures.[3] Developed samples were put in an evaporator where 2 nm Ti was evaporated as an adhesion layer followed by 35 nm of gold. Lift-off was performed by soaking in acetone for two days. Some areas of the sample were imaged with SEM to ensure successful samples, while all optical measurements were performed on areas not exposed to the electron beam.

Hyperspectral dark-field scattering spectroscopy:

A Zeiss inverted microscope (Axio observer D) was used in our home built set up.[4] Light from a halogen lamp was focused by a 1.4 NA condenser onto the sample and the scattered light was collected with a 63× 0.7 NA objective. The scattered light image was dispersed through a slit with a 100 μm width and projected onto grating, all positioned on a mechanical stage. By moving the stage and hence the spectrometer slit step by step across the scattering image, a hyperspectral image cube was formed where in the *x*-direction a spatial slice of the sample image was acquired on a 2-dimensional camera array and the spatially resolved spectrum of that slice was resolved in the *y*-direction. Following image acquisition, particles were located in the image that also contained the corresponding scattering spectra.

Photothermal imaging:

Photothermal imaging was performed on a Zeiss inverted microscope (Axio observer A1). The pump-probe geometry consists of a 532 nm laser diode (OBIS Coherent) probe and a tunable VIS/NIR pump (SuperK NKT Photonics) controlled using an acousto-optic tunable filter. The lasers were collinearly focused on the sample using a 63× 1.4 NA oil immersion objective. A 50× 0.7 NA air objective was used to collect the light transmitted through the sample. The pump laser was filtered out using a band pass filter. The transmitted probe beam was focused on a Si photoreceiver with a low noise voltage amplifier and processed with a lock-in amplifier which used the signal driving the acousto-optic tunable filter as the reference. The photothermal image was formed by scanning the sample across the focused beam using a piezo positioning stage. Absorption spectra of individual structures were obtained by recording the transmitted probe (210 µW at the sample plane) signal intensity at the centroid of a single structure and scanning the pump wavelength (500–1000 nm) with a power of ~180 µW at 700 nm using an acousto-optic tunable filter. A reference gold film next to the structures was used to correct for changes in pump power and focus as the pump wavelength was scanned as explained in a previous publication.[5]

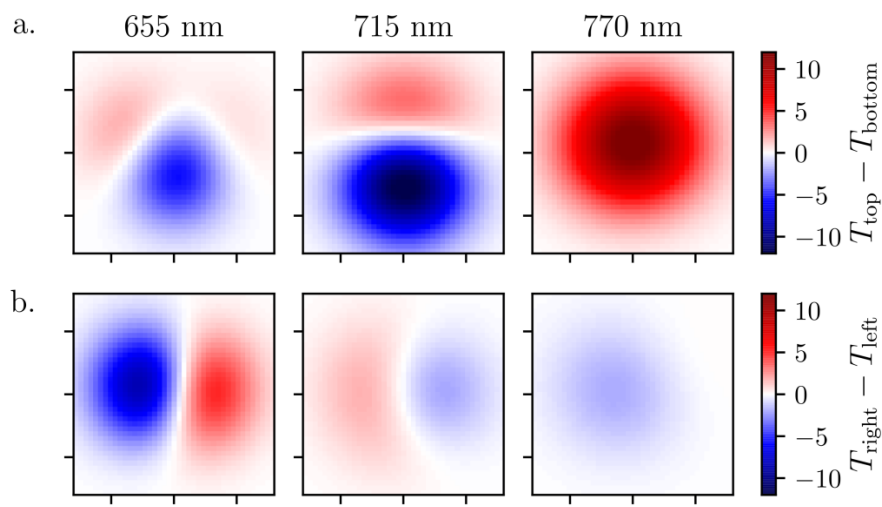

Figure S1. Temperature differences per beam position at the three normal mode wavelengths. a) The difference between the average temperature of the top two particles and the temperature of the bottom particle is shown. Each pixel corresponds to this value at a different centroid beam position. At the lowest energy mode, 770 nm, there are no beam positions where the bottom particle is hotter than the top two. At 715 nm and 655 nm, the hottest region of the system changes as a function of beam position. When the beam is centered along the bottom part of the gold nanorod trimer, the bottom particle is hotter than the top two particles. Oppositely, when the beam is in the upper portion, the top two particles become hotter than the bottom particle. However, in this case the difference between the top and bottom temperatures is smaller (fainter red versus brighter blue). b) The difference in temperature between the right and left upper particles while the bottom particle is not considered. At both the two lowest energy modes, the two temperatures are approximately the same. The highest energy mode, however, shows a strong beam induced asymmetry wherein the left particle is hotter than the right particle when the beam is in the left region. Similarly, the right particle is hotter than the left when the beam is in the right region.

**Section S3: Alternative photothermal image function**

The photothermal image expression in Eq. 1, described in Refs. 6-7 and used in our past work,[8] is based on the assumption that the lock-in measurement underlying the signal reduces to the difference in probe scattering with the pump beam on and off. Alternatively, Ref. 9 explicitly accounts for lock-in detection, but arrives at the alterative photothermal image expression

$$I^{PT} = \frac{cn}{8\pi} \iint\limits_{(\theta_d, \phi_d)} \left( [|\mathbf{E}_H(\theta,\phi)|^2 - |\mathbf{E}_R(\theta,\phi)|^2] + 2[\mathbf{E}_p(\theta,\phi) \cdot [\mathbf{E}_H(\theta,\phi) - \mathbf{E}_R(\theta,\phi)]^*] \right) d\Omega$$

where $\mathbf{E}_p(\theta,\phi)$ is the probe beam (in the absence of the target) modeled as a Gaussian beam. Careful examination of these two photothermal image functions applied to the nanorod trimer produces negligible differences for all results described in the main text. For comparison, Figure 5b,c is recreated in S3 Figure 1 where the new black dashed line indicates the FWHM ratio ($FWHM_x$/ $FWHM_y$) of the fits to the photothermal images calculated from Equation S1.

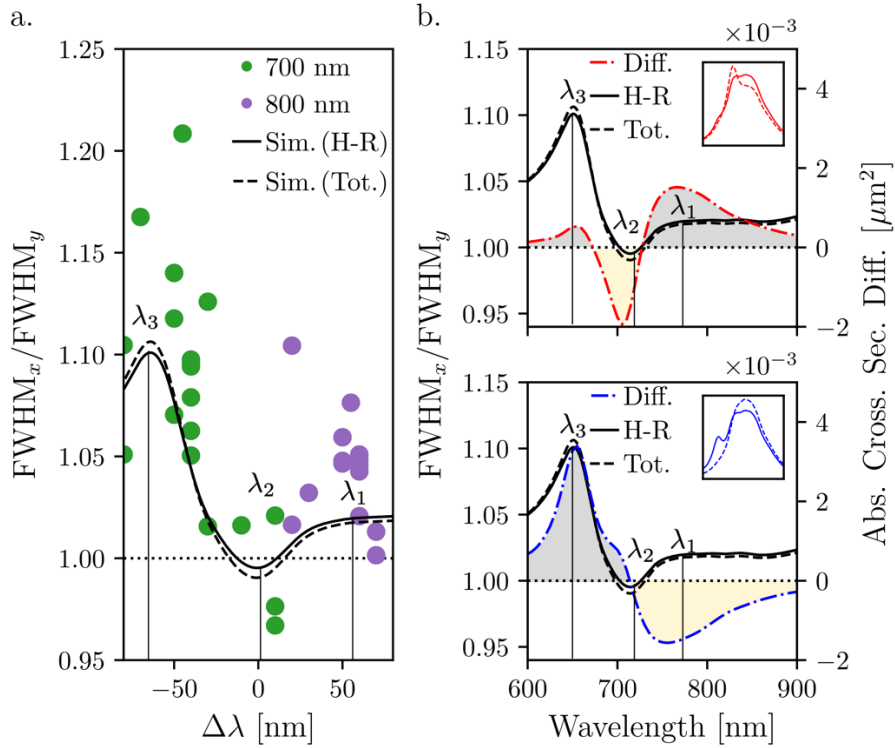

S3 Figure S1: Reconstruction of Figure 5b,c with the FWHM ratio of the new photothermal image expression overlayed in dashed black, showing near-identical behavior to that described by Eq. 1.

**Section S4: Temperature Dependent Ellipsometry of Gold**

Spectroscopic ellipsometry was conducted using a M-2000 Ellipsometer (J.A. Woollam) on a thin gold film (~27.5 nm) deposited on a glass substrate. Data were collected at five different incident angles (55, 60, 65, 70, and 75°) for three seconds each. The sample was fabricated by evaporating ~30 nm of gold onto a $SiO_2$ substrate. A piece of tape was placed on the underside of the glass substrate to eliminate out backside reflections when performing the measurement. The sample temperature was controlled using a Linkam temperature-control (THMS 600) stage with uncertainty of 0.1°C. The sample was held at the chosen temperature for ~3 min before collecting data to ensure the sample was at the proper temperature. A layer of thermal paste was also applied between the sample and the Linkam stage to reduce heat transfer resistance. The gold layer was modeled using two general oscillators, a Drude oscillator at long wavelengths and a Gaussian oscillator at short wavelengths to account for the edge of the bulk plasmon resonance of gold.

S4 Figure 1 shows the measured amplitude ratio, $\Psi$, and phase difference, $\Delta$, at 25°C (incident angle increases in the direction indicated by the arrow). The dip in $\Psi$ at higher wavelengths is due to scattering of light by the gold crystalline domains, which results in large depolarization of the incident light beam, making it difficult to properly model the data beyond ~1100 nm. However, the data between 500 and 1050 nm is sufficient for modeling the dielectric properties of gold. Data below 500 nm were also excluded due to beam instabilities and high modeling errors.

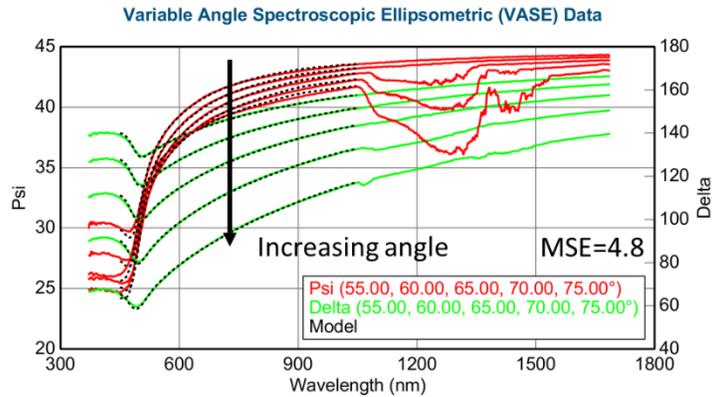

S4 Figure 1: Raw ellipsometric angle data for a 27.5 nm gold film on glass substrate as a function of incident beam angle and wavelength. The dashed lines represent fits to the data assuming the tabulated real and imaginary refractive indices for gold as supplied by the ellipsometry manufacturer.

All of the data analysis began by modeling the experimental gold film with Woollams' supplied list of tabulated real and imaginary values for the refractive index of gold (denoted $n$ and $k$, respectively). This approach typically resulted in a mean square error (MSE) of ~4–6 between the model and the data and gave a film thickness of $28.11 \pm 0.03$ nm with a roughness of $1.7 \pm 0.1$ nm at 25°C. A splining model with at least 6 nodes was then used to model the data while keeping the film thickness and roughness constant. The spline fit was also set to be Kramers-Kroning consistent, typically reducing the MSE to between 1.05 and 1.25. Note that due to the way MSE is defined, a 'perfect' fit has an MSE of 1.00.

The spline fit was then used as the basis for parametrizing the model using a series of oscillators so that the model produced from the spline can be assigned some physical meaning. Two oscillators were used to fit the data at each temperature point. The first is a Drude oscillator which is primarily used to model the optical response at longer wavelengths. This approach is typical of most metals. The second oscillator is a Gaussian function that is used to semi-empirically account for the upturn at shorter wavelengths. This is a feature that is also seen in Woollam's model for gold. While this Gaussian function could in theory be neglected for the sake of simplifying the fit, we would then have insufficient data to perform the spline fit properly so that it is Kramers-Kroning consistent. However, the MSEs of the resulting fits using the oscillator pair are still between 1.2 and 1.3, indicating a good fit.

The real and imaginary portions of the refractive index as a function of wavelength at 25ºC are shown in S4 Figure 2. The black dashed lines represent the optical properties of bulk gold that were used as the basis for the spline fit. At short wavelengths, the *n* of bulk gold tends to increase significantly. This region is generally excluded from the fit to avoid overparameterizing with added oscillators. The model results are generally related to those of the supplied values, with the only major difference being the location of the minimum of *n*. The final measured temperature dependent refractive indices for gold as a function of wavelength, by fitting the oscillator parameters at each measured temperature. These data are presented in S1 Figure 3.

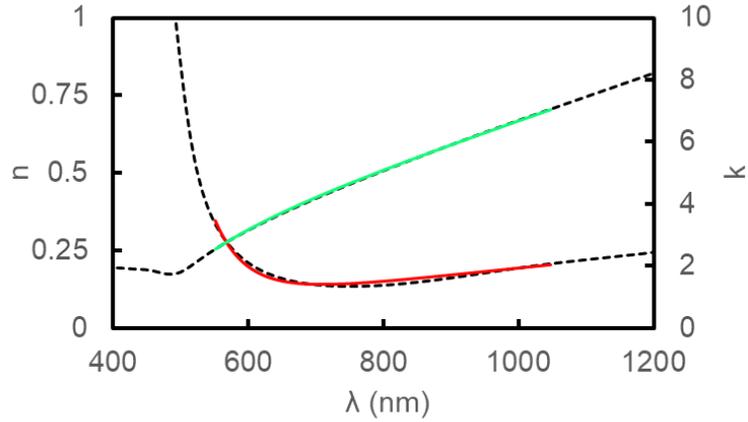

S4 Figure 2. *n* (red) and *k* (green) of the gold film as a function of wavelength. The *n* and *k* supplied by Woollam—which were used as the basis for fitting—are shown by the black dashed lines.

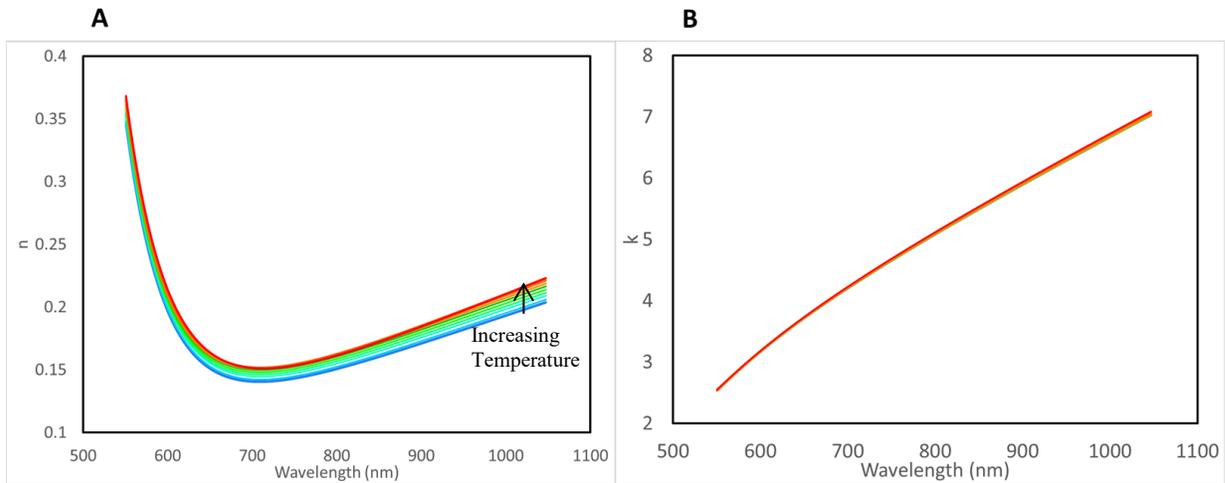

S4 Figure 3. a) Measured *n* and b) *k* of the gold film as a function of wavelength. Temperature increases blue to red.

The data in S4 Figure 3 were then fit to a Drude-Lorentz model,

$$\varepsilon(\omega) = \varepsilon_\infty - \frac{\omega_p^2}{\omega^2 + i\gamma\omega} + \frac{\omega_{pl}^2}{\omega_{0l}^2 - \omega^2 - i\gamma_l\omega}$$

and the resulting data are shown in S4 Table 1. To obtain temperature dependent refractive index data at temperatures between the ones experimentally reported, the data was interpolated using a spline fit.

S4 Table 1. Tabulated data of Drude-Lorentz parameters fit to the temperature dependent refractive index values of gold obtained from S4 Figure 3.

| $T$ [°C] | $\varepsilon_\infty$ | $\omega_p$ [eV] | $\gamma$ [eV] | $\omega_{pl}$ [eV] | $\omega_{0l}$ [eV] | $\gamma_l$ [eV] |
|---|---|---|---|---|---|---|
| 25 | 3.86588 | 8.87021 | 0.05946 | 4.23673 | 2.85633 | 0.30606 |
| 35 | 3.87538 | 8.87164 | 0.05961 | 4.24258 | 2.85602 | 0.30637 |
| 45 | 3.83424 | 8.87262 | 0.06015 | 4.29278 | 2.86494 | 0.30901 |
| 55 | 3.75522 | 8.87204 | 0.06111 | 4.34998 | 2.87144 | 0.311280 |
| 65 | 3.69099 | 8.87261 | 0.06185 | 4.40525 | 2.87925 | 0.31179 |
| 75 | 3.67691 | 8.87627 | 0.06255 | 4.41647 | 2.87859 | 0.31295 |
| 85 | 3.67729 | 8.88205 | 0.06323 | 4.42700 | 2.87989 | 0.31542 |
| 95 | 3.71633 | 8.89120 | 0.06377 | 4.41402 | 2.87818 | 0.31684 |
| 105 | 3.81036 | 8.90496 | 0.06430 | 4.35783 | 2.86824 | 0.31802 |
| 115 | 4.05619 | 8.93244 | 0.06440 | 4.20680 | 2.84488 | 0.32070 |
| 125 | 4.35593 | 8.96453 | 0.06400 | 4.03783 | 2.820030 | 0.32710 |

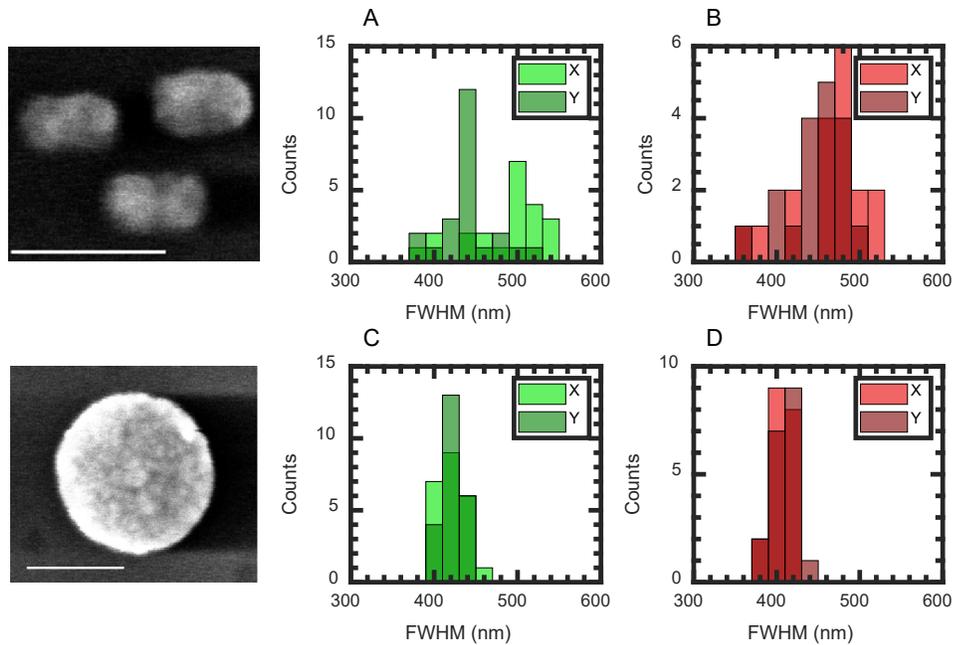

Figure S2. Absolute FWHM values of the disks and trimers. a) Trimer FWHM in *x*- and *y*-directions at 700 nm, illustrating a larger point spread function (PSF) width in the *x*-direction. b) Trimer FWHM in *x*- and *y*-directions at 800 nm indicating almost symmetric PSFs. c,d) The PSFs of the disks have the same FWHM in both directions for two excitation wavelengths of 700 nm and 800 nm, as expected for this control. The FWHM of the disk PSF is around 400 nm for both wavelengths illustrating that the PSF is independent of the pump wavelength. The 400 nm FWHM is larger than the diffraction limit of the focused probe laser (360 nm) due to convolution of the probe beam with the nanodisk size. The scale bar is 100 nm.

Table S1. Summarized FWHM values in x and y direction at two pump wavelengths of 700 nm and 800 nm. The results show a significant difference at 700 nm and a smaller difference at 800 nm. P values are the probability of x and y being identical which shows significant difference if it is less than 0.05.

|   | Mean $FWHM_{700}$ (nm) | Standard Deviation $FWHM_{700}$ (nm) | P value x vs y | Mean $FWHM_{800}$ (nm) | Standard Deviation $FWHM_{800}$ (nm) | P value x vs y |
|---|---|---|---|---|---|---|
| x | 488 | 44 | $6 \times 10^{-24}$ | 472 | 63 | 0.017 |
| y | 433 | 32 |  | 457 | 75 |  |

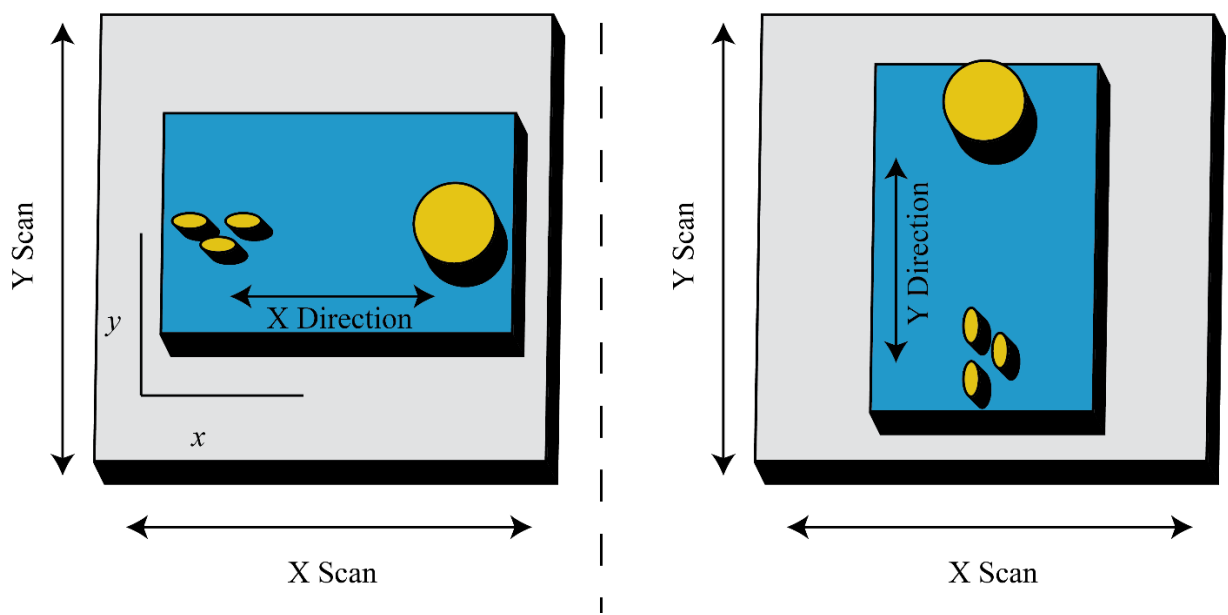

Figure S3. Schematic of scan direction and sample orientation imaging conditions. X (Y) Scan refer to the fast piezo scan direction, i.e., X-Scan is raster scanning the sample across the laser focus quickly in *x* followed by a step in *y* and repeated until an image is formed. X (Y) Direction refer to the sample orientation on the piezo scanning stage, where X Direction is when the nanorod trimer is oriented with the long axis in *x*, while Y Direction is when the nanorod trimer is oriented with the long axis in *y*. The pump polarization is always oriented parallel with the nanorod trimer long axis.

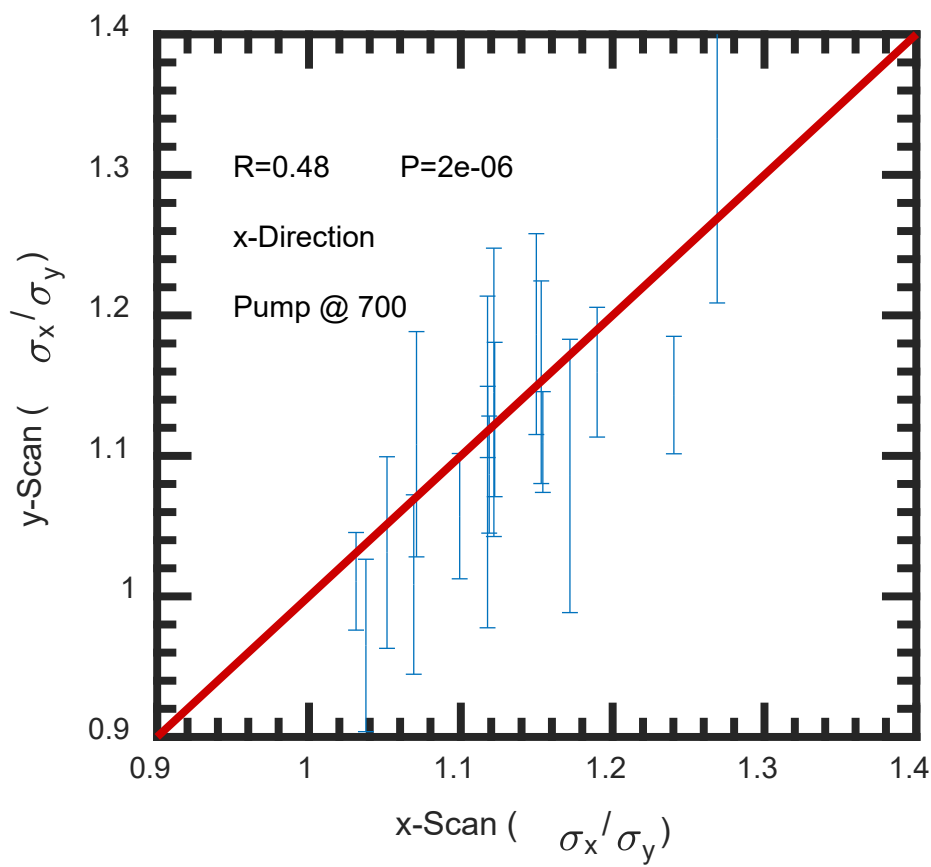

Figure S4 Correlation plot of asymmetry values using different scan directions at 700 nm. The probability for these experiments to not be correlated is small and the correlation coefficient is positive which shows changing scan direction has no effect on PSF asymmetries. P and R values are calculated by performing a Pearson correlation analysis between two data sets of measurements. R is the Pearson correlation coefficient with values ranging from -1 to +1 with (-/+)1 corresponding to a strong (negative/positive) linear relationship between the two variables. Note that R=1 does not mean a 1:1 relationship. An R of 0 indicates no linear correlation but not necessarily that the relationship is 1:0. The P value is the statistical significance that the correlation coefficient R is significantly different from 0, with $P<0.05$ corresponding to a significant difference. The red line illustrates a 1:1 relationship.

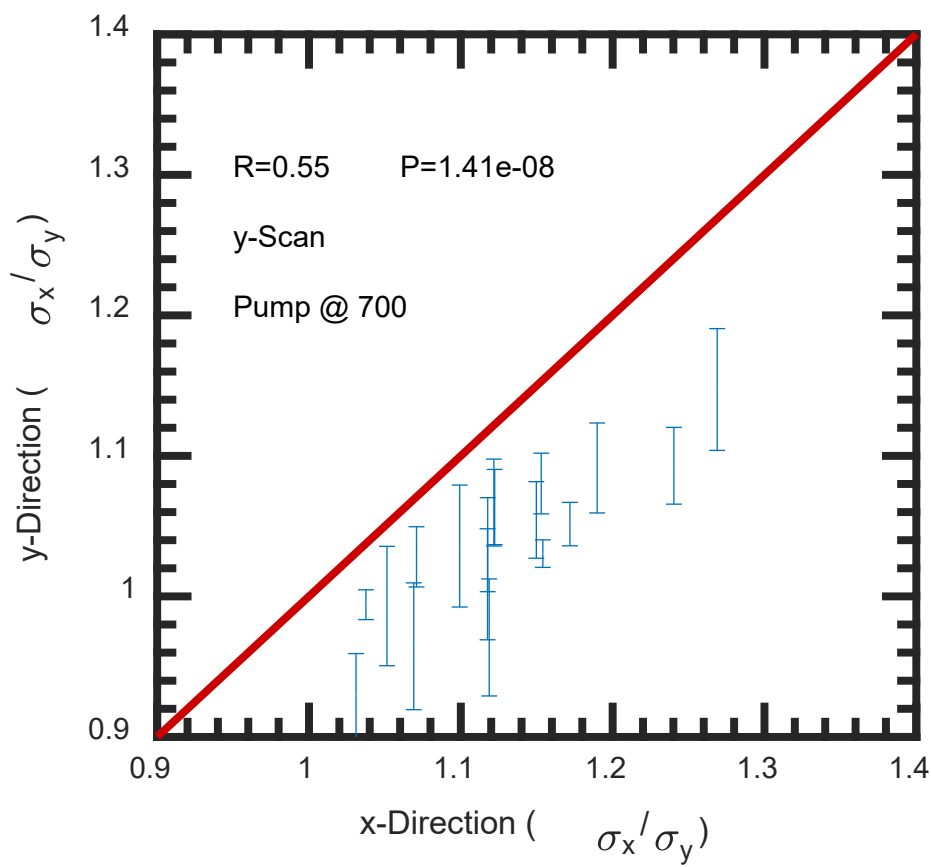

Figure S5. Correlation plot of asymmetry values with sample at different orientations at 700 nm. The probability of the two experiments not being correlated is small and the correlation coefficient is positive which shows changing sample orientation does not affect PSF asymmetries. When the sample orientation is changed the pump polarization is also changed to be able to excite the longitudinal modes on the nanorods (Figure S3).

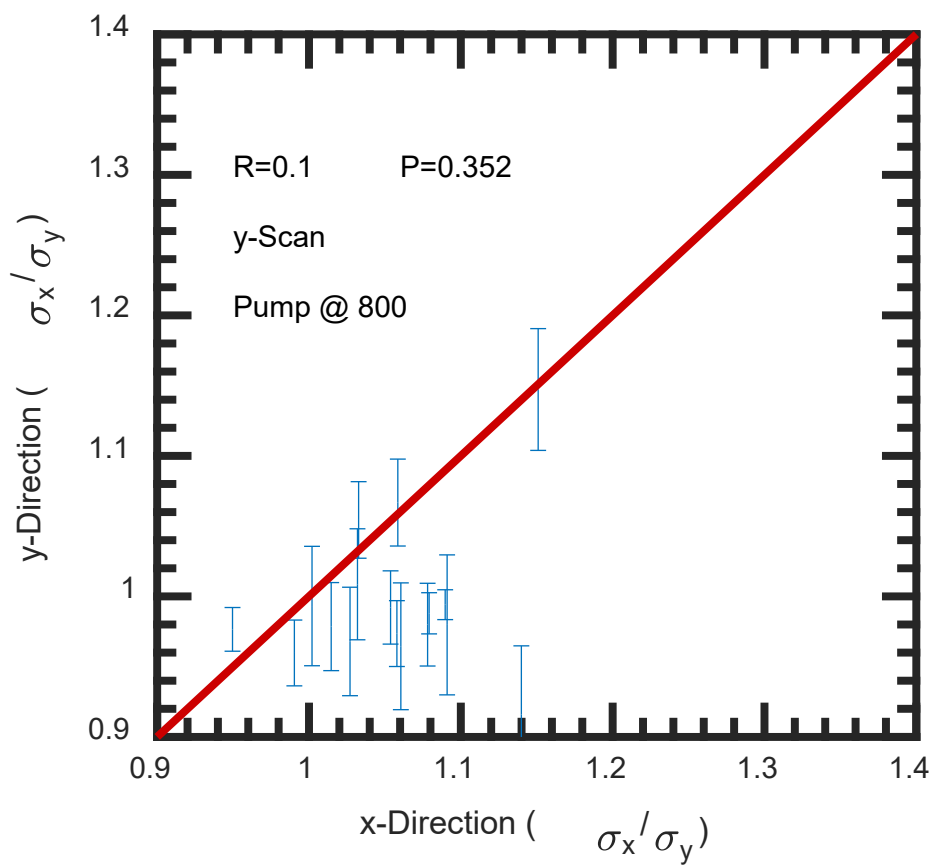

Figure S6. Correlation plot of asymmetry values using different scan direction at 800 nm. The large probability value with small correlation coefficient shows that the data are not correlated due to a smaller asymmetry in the PSF for 800 nm excitation.

S17

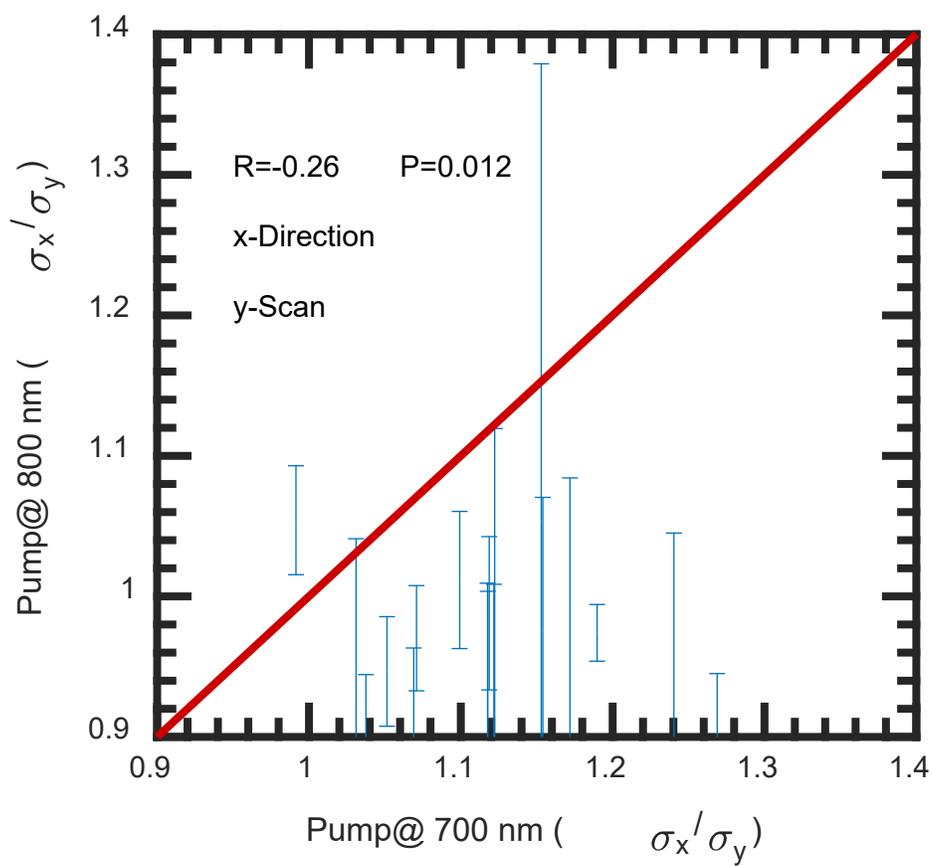

Figure S7. Correlation plot of asymmetry values at 700 nm and 800 nm. The negative correlation coefficient indicates a dependence between the measurements at these two wavelengths.